\documentclass[12pt]{article}
\usepackage{graphicx}

\newcommand\pubnumber{NuPhys2015-Duffy}
\newcommand\pubdate{\today}

\def\email{\footnote{kirsty.duffy@physics.ox.ac.uk}}

\def\Title#1{\begin{center} {\Large #1 } \end{center}}
\def\Author#1{\begin{center}{ \sc #1} \end{center}}
\def\Address#1{\begin{center}{ \it #1} \end{center}}

\newcommand\pubblock{\rightline{\begin{tabular}{l} \pubnumber\\
         \pubdate  \end{tabular}}}
\newenvironment{Abstract}{\begin{quotation}  }{\end{quotation}}
\newenvironment{Presented}{\begin{quotation} \begin{center} 
             PRESENTED AT\end{center}\bigskip 
      \begin{center}\begin{large}}{\end{large}\end{center} \end{quotation}}

\def\beq{\begin{equation}}
\def\eeq#1{\label{#1}\end{equation}}
\def\eeqn{\end{equation}}

\def\beqa{\begin{eqnarray}}
\def\eeqa#1{\label{#1}\end{eqnarray}}
\def\eeqan{\end{eqnarray}}

\let\bar=\overbar

\def\Dslash{\not{\hbox{\kern-4pt $D$}}}
\def\dslash{\not{\hbox{\kern-2pt $\del$}}}

\def\msb{{\bar{\ssstyle M \kern -1pt S}}}

\begin{document}
	
\begin{titlepage}
\pubblock

\vfill
\Title{First Antineutrino Oscillation Results from T2K}
\vfill
\Author{Kirsty Duffy\email \\ On behalf of the T2K Collaboration}
\Address{University of Oxford \\Department of Physics \\
	Denys Wilkinson Building \\ Oxford, OX1 3RH \\ United Kingdom}
\vfill
\begin{Abstract}
As limits improve on the neutrino mixing angles and mass-squared differences, the focus of T2K has shifted towards studying antineutrino oscillation. This will give an insight into CP violation (if P($\bar{\nu}_{\mu} \rightarrow \bar{\nu}_{e}) \neq \mbox{P}(\nu_{\mu} \rightarrow \nu_e)$) and CPT violation (if P($\bar{\nu}_{\mu} \rightarrow \bar{\nu}_{\mu}) \neq \mbox{P}(\nu_{\mu} \rightarrow \nu_{\mu})$) in the lepton sector. This poster summarises the most recent T2K antineutrino oscillation results, from data collected using a $\bar{\nu}_{\mu}$-enhanced neutrino beam corresponding to $4.01 \times 10^{20}$ protons on target (roughly $1/3$ of the total protons on target collected by T2K). We present world-leading measurements of $\Delta \bar{m}^2_{32}$ and $\sin^2\bar{\theta}_{23}$ and the first analysis of $\bar{\nu}_e$ appearance from T2K. Both results use a Bayesian oscillation analysis based on a Markov Chain Monte Carlo method in which data from the near detector and far detector are fit simultaneously.
\end{Abstract}
\vfill
\begin{Presented}
NuPhys2015, Prospects in Neutrino Physics \\
Barbican centre, London, UK, December 16--18, 2015
\end{Presented}
\vfill
\end{titlepage}
\def\thefootnote{\fnsymbol{footnote}}
\setcounter{footnote}{0}

\subsection*{The T2K experiment\label{sec:t2k}}

T2K is a long-baseline neutrino oscillation experiment located in Japan~\cite{T2K_NIM}, which uses the 30 GeV proton beam from the J-PARC accelerator to create a muon neutrino beam. The neutrino beam is measured by two near detectors located 280 m from the target, and a far detector, Super-Kamiokande. 
The far detector and one of the near detectors are placed 2.5$^{\circ}$ off-axis with respect to the neutrino beam, which results in a quasi-monochromatic neutrino energy spectrum that is sharply peaked around 0.6 GeV. The baseline between neutrino production and the far detector, 295 km, is carefully chosen to correspond to the first minimum in the $\nu_{\mu}$ survival probability at the peak energy.

The on-axis near detector, INGRID, an array of iron/scintillator detectors, is used to measure the beam stability, profile, and direction, and has shown that the beam direction is stable to within 0.4 mrad.
The off-axis detector, ND280, is used directly in the oscillation fits. It is made up of many subdetectors, but only the central `tracker' region is used for this analysis, comprised of two Fine-Grained Detectors (FGDs, which provide a target for neutrino interactions with excellent vertexing capabilities), and three Time Projections Chambers (TPCs, which measure the interaction products and give very good momentum resolution and particle identification). ND280 is contained in the repurposed UA1 magnet, which enables the TPC information to distinguish positive and negative charged leptons from $\overline{\nu}$ and $\nu$ interactions.

The far detector, Super-Kamiokande~\cite{SKpaper}, is a 50kton water Cherenkov detector. It has no magnetic field so cannot distinguish between neutrino and antineutrino interactions, but is capable of very good $\mu$/$e$ separation by the pattern of light from the charged lepton.

\subsection*{Oscillation analyses at T2K\label{sec:OA}}

The analysis strategy for the oscillation results presented here is similar to previous T2K results~\cite{T2K_longosc}: data samples of charged current (CC) interactions are fit at ND280 to provide a tuned prediction of the unoscillated spectrum at the far detector and its associated uncertainty. This is then compared to the data at the far detector, where $\mu$-like or $e$-like data samples are fit to estimate the oscillation parameters. The analyses presented here use data corresponding 0.43$\times 10^{20}$ protons on target (POT) in antineutrino mode plus 5.82$\times 10^{20}$ POT in neutrino mode for the near detector fit and 4.011$\times 10^{20}$ POT in antineutrino mode for the oscillation fits.
 
The predictions at both ND280 and Super-Kamiokande use the same cross-section model, so the ND280 fit can reduce the cross-section uncertainty in the Super-Kamiokande prediction by fitting parameter values in the underlying models. The (anti-)neutrino flux at Super-Kamiokande is also estimated, through correlations between the flux at the near and far detectors from theoretical models, as well as any correlations between the flux and cross-section parameters.

\begin{figure}[h]
	\begin{tabular}{p{0.5\textwidth}p{0.5\textwidth}}
		\vspace{7pt} \includegraphics[width=0.4\textwidth, height=5cm]{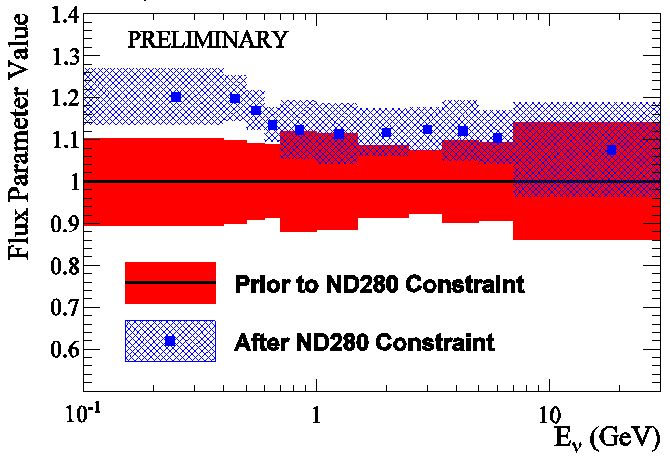} &
		\vspace{0pt} \includegraphics[width=0.4\textwidth]{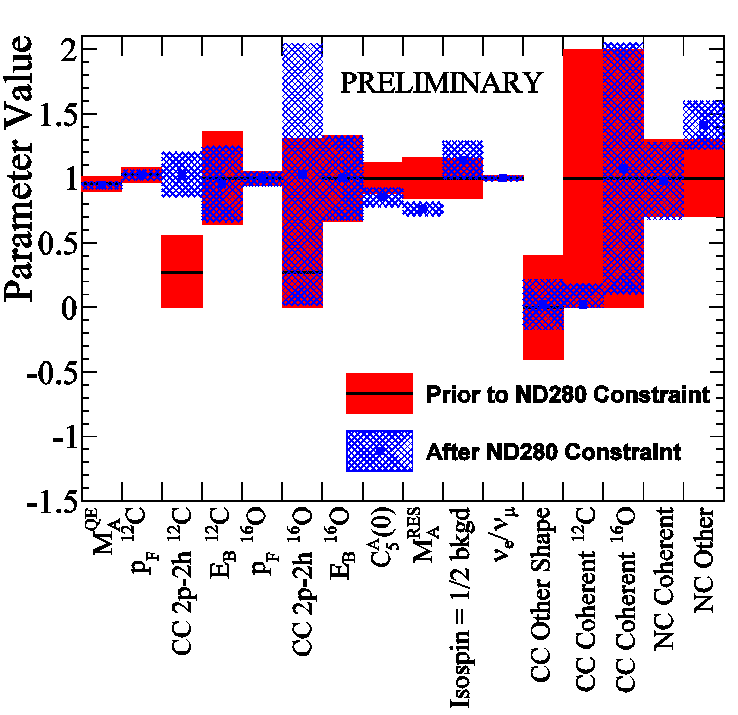}
	\end{tabular}
	\caption{\label{fig:ndfit}A subset of systematic parameters with uncertainties before and after the near detector fit. \textbf{Left:} flux parameters for $\overline{\nu}_{\mu}$ flux in the antineutrino-mode beam. \textbf{Right:} underlying parameters for the cross-section models.}
\end{figure}

Figure~\ref{fig:ndfit} shows some of the flux and cross-section parameters with their associated uncertainties before and after the near detector fit. The predicted flux at Super-Kamiokande is generally increased by the fit, although the uncertainty is decreased. Similarly, the uncertainties on the cross-section parameters to which the near detector is sensitive are generally decreased in the fit, but there is not much change to the uncertainties of the oxygen-specific parameters. 

\subsubsection*{$\overline{\nu}_e$ appearance analysis\label{sec:nuebarapp}}
	
The aim of this analysis is to look for anti-electron neutrino appearance, separately from electron neutrino appearance
. To do so, we introduce a new parameter $\beta$ which modifies the $\overline{\nu}_e$ appearance probability:
	
\begin{equation}
P(\overline{\nu}_{\mu} \rightarrow \overline{\nu}_e) = \beta \times P_{PMNS}(\overline{\nu}_{\mu} \rightarrow \overline{\nu}_e) \nonumber
\end{equation}
	
\noindent where $\beta = 1$ corresponds to $\overline{\nu}_e$ appearance in accordance with the PMNS prediction (which allows for CP violation if $\delta_{CP} \neq 0$). $\beta=0$ corresponds to no $\overline{\nu}_e$ appearance.

We report the significance for $\beta = 1$ in two ways: a p-value and a Bayes factor. The p-value relies on a test statistic: $-2\Delta \ln \mathcal{L} = -2(\ln\mathcal{L}(\beta = 1) - \ln\mathcal{L}(\beta = 0))$, where $\mathcal{L}$ is the marginal likelihood (which is integrated over all parameters other than $\beta$). This is then compared to the same test statistic calculated from an ensemble of test experiments on fake data generated with $\beta=0$, to characterise how anomalous our data are with respect to the $\beta = 0$ hypothesis. The Bayes factor is simply the likelihood ratio, and describes how much our data favours $\beta = 1$ over $\beta = 0$.

The Super-Kamiokande data and prediction are binned in either reconstructed (anti-)neutrino energy ($\overline{\nu} E_{rec}$) or momentum and angle -- with respect to the incoming neutrino direction -- of the measured lepton ($p-\theta$). Priors on the oscillation parameters are taken from the posterior of the T2K joint $\nu_{\mu}$ and $\nu_e$ fit~\cite{T2K_longosc}, which have a peak value at $\delta_{CP} \sim -\pi/2$. The current data set contains 3 events in the $e$-like sample, compared to an expected $\sim$ 1.3 events if $\beta = 0$ and $\sim$ 3.7 events if $\beta = 1$ (at oscillation parameters taken from~\cite{T2K_longosc}).
	
Table~\ref{tab:nuebarresults} shows the results of this analysis: the p-value is greater than 15\% and the Bayes factor is around 1 -- neither of these show strong enough evidence to support $\beta=1$ over $\beta = 0$, so with the current data set we cannot conclude that we have observed $\overline{\nu}_e$ appearance.
	
\begin{table}[h]
	\centering
	\small
	\begin{tabular}{cccc}
		\hline
		Super-Kamiokande binning & $-2\Delta \ln \mathcal{L}$ & p-value & B$_{10}$ \\
		\hline
		$\overline{\nu} E_{rec}$ & 0.16 & 0.16 & 1.1 \\
		Lepton $p-\theta$ & -1.16 & 0.34 & 0.6 \\
		\hline
	\end{tabular}
	\caption{\label{tab:nuebarresults}Test statistics, p-values, and Bayes factors from T2K $\overline{\nu}_e$ appearance analysis}
\end{table}

\subsubsection*{$\overline{\nu}_{\mu}$ disappearance analysis\label{sec:numubardisapp}}
		
This analysis uses the antineutrino-mode data to measure the antineutrino oscillation parameters, so CPT invariance is not assumed. We fit the oscillation parameters that dominate $\overline{\nu}_{\mu}$ disappearance, $\sin^2\overline{\theta}_{23}$ and $\Delta \overline m^2_{32}$, and fix all other antineutrino and neutrino oscillation parameters to values taken from the results of previous T2K fits~\cite{T2K_longosc} or the 2014 edition of the PDG~\cite{PDG2014}. The Super-Kamiokande data is binned only in reconstructed (anti-)neutrino energy, and contains 34 events in the $\mu$-like sample from the antineutrino-mode beam. 

\begin{figure}[h]
	\includegraphics[width=0.49\textwidth]{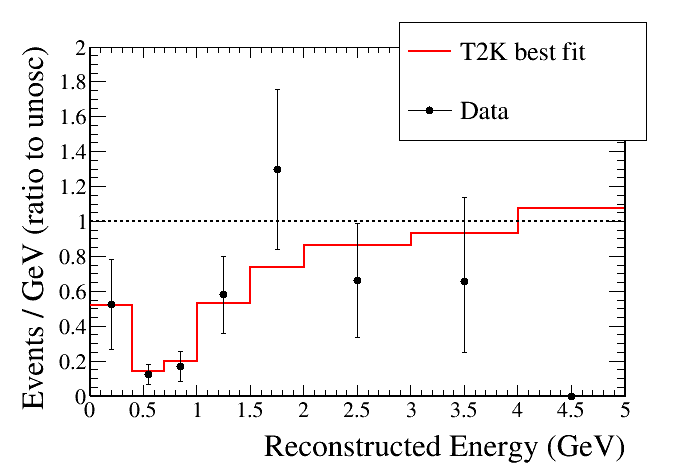}
	\includegraphics[width=0.49\textwidth]{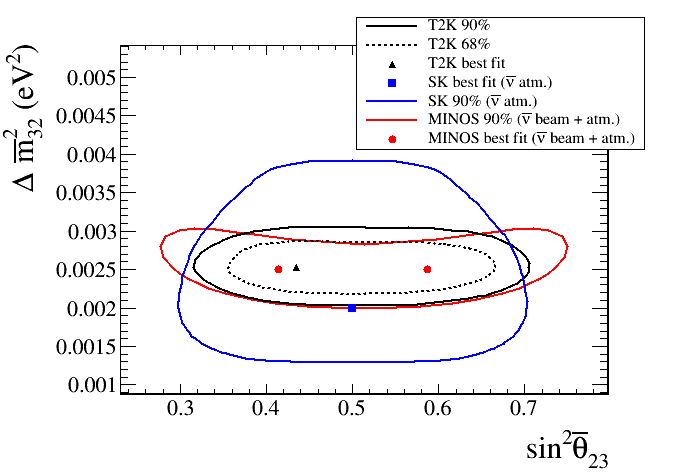}
	\caption{\label{fig:numubar:erec}\textbf{Left:} Ratio of best-fit energy spectrum and data to prediction without oscillations. \textbf{Right:} 68\% and 90\% credible intervals in $\sin^2\overline{\theta}_{23}$--$\Delta \overline m^2_{32}$ overlaid with contours from similar analyses by MINOS~\cite{MINOS_numubar} and Super-Kamiokande~\cite{SK_numubar}.}
\end{figure}

The left-hand plot in figure~\ref{fig:numubar:erec} shows the best-fit spectrum and data as a ratio to the unoscillated prediction, with the characteristic `oscillation dip' that is clear evidence of $\overline{\nu}_{\mu}$ disappearance. The right-hand side of figure~\ref{fig:numubar:erec} shows the 68\% and 90\% credible interval contours in $\sin^2\overline{\theta}_{23}$--$\Delta \overline m^2_{32}$  compared to the 90\% contours from $\overline{\nu}_{\mu}$ disappearance analyses in MINOS (using antineutrino-mode beam and atmospheric data)~\cite{MINOS_numubar} and Super-Kamiokande (using atmospheric neutrino data only)~\cite{SK_numubar}. The MINOS contour was originally presented in terms of $\sin^22\overline{\theta}_{23}$ and had to be unfolded (hence the two best-fit points), but the results from all three experiments are in agreement.

The results of this analysis are also in agreement with previous T2K neutrino-mode measurements: we see no evidence for CPT violation. 
The best-fit parameter estimates are:

\begin{table}[!h]
	\centering
	\begin{tabular}{ccc}
		$\sin^2\overline{\theta}_{23} = 0.46^{+0.14}_{-0.06}$  & & 
			$\Delta \overline{m}^2_{32} = 2.50^{+0.3}_{-0.2}\times 10^{-3} \mbox{ eV}^2$ \\
	\end{tabular}
\end{table}

These and the 2D contours are consistent with the results presented in~\cite{numubarpaper}, which were calculated using a hybrid frequentist-Bayesian method (as opposed to the fully-Bayesian method presented here).
		
\subsection*{Summary and future prospects}
		
We have presented here the first T2K results based on antineutrino data. A search for $\overline{\nu}_e$ appearance was inconclusive with the current data set, and a measurement of $\sin^2\overline{\theta}_{23}$ and $\Delta \overline m^2_{32}$ has been made from $\overline{\nu}_{\mu}$ disappearance at T2K, which is in agreement with T2K neutrino-mode fits and antineutrino results published by MINOS and Super-Kamiokande. 
	T2K continues to run with an antineutrino beam which will provide additional data to improve both measurements.


\begin{thebibliography}{99}
%
%
%
%
\bibitem{T2K_NIM}
K. Abe et al.
, Nucl. Instr. Meth. Phys. Res. A {\bf 659}, 106-135 (2011)
%
\bibitem{SKpaper}
S. Fukuda et al.
, Nucl. Instr. Meth. Phys. Res. A {\bf 501}, 418-462 (2003)
%
\bibitem{T2K_longosc}
K. Abe et al.
, Phys Rev. D {\bf 91}, 072010 (2015)
%
%
\bibitem{PDG2014}
K. A. Olive et al.
, Chin. Phys. {\bf C38}, 090001 (2014)
%
\bibitem{MINOS_numubar}
P. Adamson et al.
, Phys. Rev. Lett. {\bf 110}, 251801 (2013)
%
\bibitem{SK_numubar}
K. Abe et al.
, Phys. Rev. Lett. {\bf 107}, 241801 (2011)
%
\bibitem{numubarpaper}
K. Abe et al.
, accepted into Phys. Rev. Lett., arXiv:1512.02495 [hep-ex] (2015)
%
\end{thebibliography}
\end{document}